\documentclass[preprint,aps]{revtex4}
\usepackage{epsf}
\usepackage{graphicx} 
\usepackage{color}

\begin{document}

\title
{
High-Order Coupled Cluster Method (CCM) Calculations for Quantum Magnets with
Valence-Bond  Ground States
}


\author{D.J.J. Farnell$^1$, J. Richter$^2$, R. Zinke$^2$, and R.F. Bishop$^3$}


\affiliation{$^1$Academic Department of Radiation Oncology, Division of Cancer Studies, Faculty of Medical and Human Science,  University of Manchester, c/o Christie Hospital NHS Foundation Trust, Manchester M20 4BX, United Kingdom}
\affiliation{$^2$Institut f\"ur Theoretische Physik, Otto-von-Guericke Universit\"at
Magdeburg, P.O.B. 4120, 39016 Magdeburg, Germany}
\affiliation{$^3$School of Physics and Astronomy, Schuster Building, The University of Manchester, Manchester M13 9PL, United Kingdom}

\date{\today}

\begin{abstract}
In this article, we prove that exact representations of dimer and 
plaquette valence-bond ket ground states  for 
quantum Heisenberg antiferromagnets may be formed 
via the usual coupled cluster method (CCM) 
from  independent-spin product (e.g. N\'eel) model states.
We show that we are able to provide good results for both the ground-state energy and the 
sublattice magnetization for dimer and plaquette valence-bond phases within the CCM. 
As a first example, we investigate the spin-half $J_1$--$J_2$ model for the linear chain, 
and we show that we are able to reproduce 
exactly the dimerized ground (ket) state at $J_2/J_1=0.5$.
The dimerized phase is stable over a range of 
values for $J_2/J_1$ around 0.5, and results for 
the ground-state energies are in good agreement with the results of 
exact diagonalizations  of finite-length chains in this regime. 
We present evidence of symmetry breaking by considering the ket- and 
bra-state correlation coefficients as a 
function of $J_2/J_1$. A radical change is also observed in the behavior of the
CCM sublattice magnetization as we enter the dimerized 
phase. We then consider the Shastry-Sutherland model and demonstrate 
that the CCM can span the correct ground states in both the 
N\'eel and the dimerized phases. Once again, very good results for the ground-state energies are obtained. 
We find CCM critical points of the bra-state equations that are in 
agreement with the known  phase transition point for this model.  The results for the 
sublattice magnetization remain near to the ``true'' value of 
zero over much of the dimerized regime, 
%(e.g., $|M|<10^{-2$ at the LSUB6 
%level of approximation for $J_2/J_1>2$), 
although they diverge exactly 
at the critical point.  Finally, we consider a spin-half system with 
nearest-neighbor bonds for an underlying lattice corresponding to 
the magnetic material CaV$_4$O$_9$ (CAVO). 
We show that we are able to provide excellent results for the 
ground-state energy in each of the plaquette-ordered, N\'eel-ordered, and dimerized regimes of this model. 
The exact plaquette and dimer ground states are reproduced by the CCM ket state in their relevant limits. 
Furthermore, we estimate the range 
over which the N\'eel order is stable, and we find  the  CCM result is in reasonable 
agreement with the results obtained by other methods. 
Our new approach has the dual advantages that it is simple to implement 
and that existing CCM 
codes for  independent-spin product  model states 
may be used from the outset. Furthermore, it also greatly extends the range of applicability 
to which the CCM may be applied. We believe that 
the CCM now provides an excellent choice of method for the study of systems with 
valence-bond quantum ground states.
\end{abstract}

\maketitle

\section{Introduction}
\label{intro}
Lattice quantum spin models not only provide useful models of many 
physically realizable magnetic systems but also serve as prototypical models of strongly interacting 
quantum many-body systems. Indeed, the basic models of quantum magnets are given by lattice spin models that often 
display rich quantum phase transitions between ground states of different 
order as some control parameter is varied.  
Their collective behavior is extremely complex due to the 
presence of strong quantum effects. 
Furthermore, the underlying crystallographic lattices for these materials may exhibit complex 
symmetries. 
Their rich phase diagrams include 
exotic phases of novel quantum order due to the strong interplay between competing 
interactions and large quantum fluctuations. 
For all of these reasons they have naturally provided an excellent 
test-bed where the various methods of quantum many-body theory can 
be applied and further refined.  

Of particular interest is the formation of dimer- and plaquette-ordered singlet 
ground states (so-called valence-bond crystal (VBC) states) 
in quantum spin systems.
Often, the formation of enhanced dimer or plaquette correlations 
is driven by frustration, which 
can increase quantum fluctuations and which may 
result in such gapped 
rotationally-invariant quantum paramagnetic states
 \cite{mg1,mg2,mg3,mg4,mg5,mg6,aligia00,shastry,shastry2,shastry3,shastry4,shastry5,shastry6,shastry7,shastry8,shastry9,shastry10,shastry11,shastry12,shastry13,shastry14,shastry15,star,star07}.
Usually, VBC states are complicated quantum many-body states, see, e.g., the Heisenberg
antiferromagnet on the star
lattice \cite{shastry2,star,star07}.  
However, for certain systems the VBC states are simple exact  product 
eigenstates of the underlying Heisenberg interaction Hamiltonian. Examples for the appearance 
of such exact VBC product eigenstates are the 
spin-half $J_1$--$J_2$ model on the linear chain 
 \cite{mg1,mg2,mg3,mg4,mg5,mg6,aligia00}  at the point $J_2/J_1=0.5$ (the 
 so-called Majumdar-Ghosh
point) and the Shastry-Sutherland model \cite{shastry,shastry2,shastry3,shastry4,shastry5,shastry6,shastry7,shastry8,shastry9,shastry10,shastry11,shastry12,shastry13,shastry14,shastry15}.
Furthermore, it is often useful to distinguish between VBC phases that have the same
translational symmetry as the spin Hamiltonian 
and those that  spontaneously break the symmetry of 
the underlying spin lattice. Examples of the former case are the
Shastry-Sutherland model and the Heisenberg antiferromagnet on the star
lattice, whereas the $J_1$--$J_2$ model on the linear chain
is an example of spontaneous symmetry breaking.

Another mechanism for the formation of non-magnetic 
dimer or plaquette VBC ground
states that does not involve frustration
is the competition between non-equivalent antiferromagnetic 
nearest-neighbor  bonds. This may  lead to the
 formation of local singlets of two (or four) 
coupled spins if the strengths of the non-equivalent  bonds differ
sufficiently \cite{shastry2,cavo1,cavo2,cavo2a,cavo3,cavo4,cavo5,cavo6,singh,ivanov,ccm17,tomczak,ccm24}. 
By contrast to frustration, which yields
competition in quantum as well as in classical spin systems, this type of
competition is present only in quantum systems. 
The symmetry of
the ground state follows the symmetry of the Hamiltonian in such cases.
An example of this is given by the 
nearest-neighbor models for the magnetic 
material CaV$_4$O$_9$ of Ref.  \cite{cavo1,cavo2,cavo2a,cavo3,cavo4,cavo5,cavo6} 
in which one finds two non-equivalent antiferromagnetic 
nearest-neighbor bonds $J$ and $J'$ 
belonging to dimers ($J'$) or to 
four-spin plaquettes ($J$). 
If these non-equivalent nearest-neighbor bonds 
are different in strength then local singlet formation may destroy N\'eel
long-range order.  
As $J'/J$ becomes large, local singlets are formed on dimers. 
Alternatively, local four-spin singlets on the plaquettes are formed for strong plaquette bonds $J$.
Another example studied in the literature is the so-called 
$J$--$J'$ model on the square lattice, i.e., a model with a regular distribution
of two different nearest-neighbor bonds on the square
lattice \cite{singh,ivanov,ccm17,tomczak,ccm24,wenzel08}. In both cases, the formation of
local singlets suppresses the magnetic N\'eel long-range order. However, the
VBC ground state is now a complex many-body state; with a simple product
VBC state appearing only in the limits $J'/J \to \infty$ and 
$J/J' \to \infty$.      

In this article we focus on the application of the
coupled cluster method (CCM) to quantum Heisenberg antiferromagnets having
VBC ground-state phases. 
The CCM  
 \cite{shastry15,ccm17,ccm24,refc1,refc2,refc3,refc4,refc5,refc6,refc7,refc8,refc9,ccm1,ccm2,ccm999,ccm3,ccm4,ccm5,ccm6,ccm7,ccm8,ccm9,ccm10,ccm11,ccm12,ccm13,ccm13a,ccm14,ccm15,ccm16,ccm18,ccm19,ccm19a,ccm20,ccm21,ccm22,ccm23,ccm,ccm26,ccm27,ccm27a,ccm28,ccm29,ccm30,ccm31,ccm32,ccm33,ccm34,ccm35} 
is one of the most powerful 
and most widely applied techniques of modern-day microscopic 
quantum many-body theory. The CCM allows the inclusion of multi-particle correlations into the 
ground- and excited-state wave functions in a controlled and systematic manner. 
It has been applied to a great variety of different lattice quantum spin systems 
with great success.  In particular, it has been used successfully with 
 model (or reference) states built
by independent-spin product states for which
the choice of state for the spin on each site
is formally independent of the choice of all others. Often for these 
independent-spin product model states the use of collinear states, such as 
the N\'eel state, is possible where all spins 
are aligned parallel or antiparallel 
to one axis (e.g., the $z$-axis), see, Refs.~\cite{ccm1,ccm2,ccm999,ccm3,ccm4,ccm5,ccm8,ccm9,ccm10,ccm11,ccm12,ccm13,ccm13a,ccm14,ccm15,ccm16,ccm18,ccm19,ccm20,ccm22,ccm26,ccm27a,ccm28,ccm29,ccm30,ccm31,ccm32,ccm33,ccm34,ccm35}. 
However, noncollinear 
(e.g. spiral) model states can be favorable for certain 
frustrated spin systems
\cite{ccm7,shastry15,ccm17,ccm19a,ccm21,ccm23,ccm27,ccm29}.
Multi-spin correlations are then included systematically on top of the independent-spin product model states.

The CCM for independent-spin product model states may be applied 
to high orders by using a 
computational implementation described in Refs.
 \cite{ccm8,ccm12,ccm15,ccm20,ccm23,ccm}. In particular, it may be applied to lattices of 
complex crystallographic symmetry. Furthermore, it is not constrained to systems with spin quantum
number $s=1/2$.

In previous work, non-classical VBC ordering 
has also been considered using the CCM by employing 
directly valence-bond model states, 
i.e. two- or four-spin singlet product states \cite{ccm6}. 
However, this earlier approach involves the direct use of products of 
localized states (e.g., two-spin dimers or four-spin plaquettes) in the model state. Hence, this 
approach requires that a whole new matrix-operator formalism be 
created for each new problem. Also, the Hamiltonian and CCM 
ket- and bra-state operators must be written in terms of this new matrix algebra. The CCM equations may be derived and solved once the commutation relationships between the operators have been established. 
Although formally straightforward, this process can be tedious and time-consuming. 
Furthermore, the existing high-order CCM formalism and codes also need to be amended extensively for each separate model considered.

In this article we use a quite different way to describe VBC states.
Starting directly from  collinear 
independent-spin product model states, 
we discuss how we can form 
exact local dimer or plaquette ground states within the CCM.
This approach has the advantages of being conceptually simple and 
thus also of being easy  to implement. 
Furthermore, we can use directly the existing high-order CCM formalism, 
computer codes, 
and extrapolation schemes. After first describing the new methodology, 
we apply the method to specific spin 
models considered in the literature that exhibit 
dimer and plaquette ground states. By comparison with existing results, 
we demonstrate that the CCM provides  good 
results for the ground-state properties of these systems. 
We then conclude the article by considering the implications 
of our results.

\section{Method}
\label{sec:2}

%We begin our explanation of the CCM method for valence-bond ground states by describing firstly 
%the underlying CCM formalism for the ground state. We then explain how dimerized and plaquette states may be formed exactly 
%using the CCM from  
%independent-spin product model states on crystallographic lattices.

\subsection{CCM Formalism}\label{ccm_f}
\label{sec:21}

The exact ket and bra ground-state energy eigenvectors, $|\Psi\rangle$ and $\langle\tilde{\Psi}|$, of a 
general many-body system described by a Hamiltonian $H$, 
%%%%%%%%%%%%%%%%%%%%%%%%%%%%%%%%%%
\begin{equation} 
H |\Psi\rangle = E_g |\Psi\rangle
\;; 
\;\;\;  
\langle\tilde{\Psi}| H = E_g \langle\tilde{\Psi}| 
\;, 
\label{eq1} 
\end{equation} 
%%%%%%%%%%%%%%%%%%%%%%%%%%%%%%%%%%  
are parametrized within the normal (NCCM) version of the single-reference CCM as follows:   
%%%%%%%%%%%%%%%%%% 
\begin{eqnarray} 
|\Psi\rangle = {\rm e}^S |\Phi\rangle \; &;&  
\;\;\; S=\sum_{I \neq 0} {\cal S}_I C_I^{+}  \nonumber \; , \\ 
\langle\tilde{\Psi}| = \langle\Phi| \tilde{S} {\rm e}^{-S} \; &;& 
\;\;\; \tilde{S} =1 + \sum_{I \neq 0} \tilde{{\cal S}}_I C_I^{-} \; .  
\label{eq2} 
\end{eqnarray} 
%%%%%%%%%%%%%%%%%% 
The normalized single model or reference state $|\Phi\rangle$ is defined 
with respect to a suitable set of (mutually commuting) many-spin creation operators $\{C_I^{+}\}$.
%(An example of a typical model state is given by $|\Phi\rangle = | \cdots \downarrow \downarrow 
 %\downarrow \downarrow \downarrow \downarrow \cdots \rangle$.)  
%For ease and generality of discussion the model state here is taken to 
%be a state in which all spins point in the downwards $z$-direction. Clearly,
%{\it any} model state can be so chosen by an appropriate choice of local
%spin coordinates on each site, and hence this choice is purely one of convention.
We note that the model states are generally related to the classical 
ground states of the lattice spin system. These states are 
products of single-spin eigenstates of some appropriately 
defined operator 
$s_i^z$ whose direction in a set of global axes can differ from site to site, 
e.g., a typical such states is:
$
\nonumber
|\Phi\rangle =~\cdots ~ \otimes|\uparrow\rangle \otimes |\downarrow 
\rangle \otimes|\uparrow  
\rangle \otimes |\downarrow 
\rangle \otimes 
\cdots$ (N\'eel state). 
However, we remark also that planar model states or 
spiral model states in the global spin coordinate 
axes 
($|\Phi\rangle = ~\cdots ~\otimes | \rightarrow 
\rangle \otimes |\leftarrow 
\rangle \otimes |\rightarrow  
\rangle \otimes |\leftarrow 
\rangle \otimes  
\cdots$ and $|\Phi\rangle =~\cdots \otimes |\uparrow 
\rangle \otimes | \nearrow 
\rangle \otimes | \rightarrow  
\rangle \otimes | \searrow 
\rangle \otimes | \downarrow 
\rangle \otimes | \swarrow 
\rangle \otimes \cdots$, respectively) may also be considered.
In order to make the CCM calculations easier to carry out in practice, 
we generally rotate the local axes of the spins so that they all appear 
notationally to point in the downwards $z$-direction. The model state is 
then given by:  $|\Phi\rangle=~\cdots ~\otimes| \downarrow
\rangle \otimes|\downarrow\rangle \otimes|\downarrow\rangle \otimes|
\downarrow\rangle \otimes\cdots~$.
The interested reader is referred to 
Refs.~\cite{ccm7,shastry15,ccm17,ccm19a,ccm21,ccm23,ccm27,ccm29} for more details about 
spiral model states 
and Ref. \cite{ccm9,ccm32} for more details about planar model states.

The operators $C_I^+ \equiv (C_I^-)^{\dag}$, with $C_0^+ \equiv 1$, have the property
that $\langle \Phi | C_I^+ = 0 = C_I^- | \Phi \rangle ~ \forall ~ I \ne 0$. They form a 
complete set of multi-spin creation operators with respect to the model state 
$| \Phi \rangle$. Thus, the creation operators are 
represented simply as a product of spin-raising operators 
$s_k^+ \equiv s_k^x + {\rm i} s_k^y$ over the set of lattice
sites $\{ k \}$ after rotation of the local frames such that all spins
appear to point downwards, as described above.
The creation operators are now given by $C_I^+ \equiv s_{i_1}^+ s_{i_2}^+ \cdots s_{i_l}^+ $.  
We note that the definitions of Eq. (\ref{eq2}) imply the normalization
$\langle \tilde \Psi | \Psi \rangle =  \langle \Phi | \Phi \rangle =  1$.

The determination of the correlation coefficients $\{ {\cal S}_I, 
\tilde{{\cal S}}_I \}$ is achieved 
%by taking appropriate projections 
%onto the ground-state 
%Schr\"odinger equations of Eq. (\ref{eq1}). Equivalently, they may be 
%determined variationally 
by requiring the ground-state energy expectation 
functional $\bar{H} ( \{ {\cal S}_I, \tilde{{\cal S}}_I\} )\equiv \langle \tilde{\Psi} | H | \Psi\rangle=
\langle\Phi | \tilde{S} {\rm e}^{-S} H {\rm e}^S | \Phi\rangle$ to be 
stationary with respect to variations in each of 
the (independent) variables of the full set. We thereby derive 
the following coupled set of equations: 
%%%%%%%%%%%%%%%%% 
\begin{eqnarray} 
\delta{\bar{H}} / \delta{\tilde{{\cal S}}_I} =0 & \Rightarrow &   
\langle\Phi|C_I^{-} {\rm e}^{-S} H {\rm e}^S|\Phi\rangle = 0 ,  \;\; 
\forall I \neq 0 \;\; ; \label{eq7} \\ 
\delta{\bar{H}} / \delta{{\cal S}_I} =0 & \Rightarrow & 
\langle\Phi|\tilde{S} {\rm e}^{-S} [H,C_I^{+}] {\rm e}^S|\Phi\rangle 
= 0 , \;\; \forall I \neq 0 \;\; . \label{eq8}
\end{eqnarray}  
%%%%%%%%%%%%%%%% 
Equation (\ref{eq7}) also shows that the ground-state energy at the stationary 
point has the simple form 
%%%%%%%%%%%%%%%%
\begin{eqnarray}
E_g &\equiv& \langle \tilde \Psi | H | \Psi \rangle = \langle \Phi | \tilde S e^{-S} H e^S | \Phi \rangle
\nonumber \\  
         &=&  
         \langle \Phi | (1 + \sum_{I \neq 0} \tilde{{\cal S}}_I C_I^{-}) e^{-S} H e^S | \Phi \rangle
  \nonumber \\
         &=&  
        \langle \Phi | e^{-S} H e^S | \Phi \rangle + 
             \sum_{I \neq 0} \tilde{{\cal S}}_I  \langle \Phi | C_I^{-} e^{-S} H e^S | \Phi \rangle
 \nonumber \\
 \Rightarrow E_g        &=&  
        \langle \Phi | e^{-S} H e^S | \Phi \rangle \;\; . \label{eq9} 
\end{eqnarray}
%%%%%%%%%%%%%%%% 
We see that the expectation value of the ground-state energy of Eq.  (\ref{eq9}) 
contains terms in $S$ only and so it also contains ket-state correlation coefficients
only. Generally, however, we need to use both the bra and ket states to 
find a ground-state expectation value. Indeed, the ground-state energy 
is the only special case that requires just the ket-state alone.
We note also that this (bi-)variational formulation does {\it not} lead to 
an upper bound for $E_g$ when the summations for $S$ and $\tilde{S}$ 
in Eq. (\ref{eq2}) are truncated, due to the lack of exact hermiticity 
when such approximations are made. However, it is also important 
to realize that the Hellmann-Feynman theorem {\it is} preserved 
in all such approximations \cite{refc9}.

We note that any practical calculation requires an approximation 
to be made for both $S$ and $\tilde{S}$. The three 
most common schemes are:  (1) the SUB$n$ scheme, 
in which all correlations 
involving only $n$ or fewer spins are retained; (2) the SUB$n$-$m$  
sub-approximation, in which all $n$-spin-flip clusters 
spanning a range of no more than $m$ adjacent lattice 
sites are retained;  and (3) the ``localized"  LSUB$m$ scheme, in which all 
multi-spin correlations over distinct locales on the 
lattice defined by $m$ or fewer contiguous sites are 
retained. 

%The method of then solving the ``CCM problem" for the bra and ket states encapsulated in Eqs. (\ref{eq7}) and (\ref{eq8}) for lattice quantum spin systems at zero temperature has been discussed extensively 
%elsewhere
 %\cite{shastry15,ccm17,ccm24,ccm1,ccm2,ccm999,ccm3,ccm4,ccm5,ccm7,ccm8,ccm9,ccm10,ccm11,ccm12,ccm13,ccm13a,ccm14,ccm15,ccm16,ccm18,ccm19,ccm19a,ccm20,ccm21,ccm22,ccm23,ccm,ccm26,ccm27,ccm27a,ccm28,ccm29,ccm30,ccm31,ccm32,ccm33,ccm34,ccm35}. 
% The manner in which the CCM equations at the LSUB$m$ level of approximation are 
%solved to high orders in the truncation index $m$ via computational 
%methods is discussed in Refs. 
 %\cite{ccm8,ccm12,ccm15,ccm20,ccm23,ccm}. 
%The sublattice magnetization in the unrotated spin frames is given 
%by $M = - \frac 1N \sum_{i=1}^N \langle \tilde \Psi \mid (-1)^i s_i^z 
%\mid \Psi \rangle$. The factor of $(-1)^i$ is removed 
%after the local frames of references for the spins have been 
%rotated, as is common for the model states used here.
%Hence, 

We use here an order parameter $M \equiv - \langle \tilde \Psi \mid s^z \mid
\Psi \rangle$ that is thus defined to be the negative of the ground-state expectation value of the operator
$s^z \equivÊ \frac {1}{N} \sum_{i=1}^N s_i^z$.Ê Here $N \rightarrow \infty$ is the total number of spins on the lattice and, very importantly, each operator $s_i^z$ is defined with respect to the {\em local rotated} spin axes on lattice site $i$, which are themselves defined by the choice of particular model state $| \Phi \rangle$, as explained previously.Ê Clearly, by definition, for the original model state, the order parameter $M$ is simply $- \langle \Phi \mid \ s^z \mid \Phi \rangle = \frac {1}{2}$.Ê Quantum correlations (or fluctuations) in the exact ground state will thus have the effect of decreasing $M$ from this maximal value.Ê For the sake of ease of use and clarity we shall henceforth refer to the order parameter $M$ as the sublattice magnetization, although this terminology is strictly only appropriate for antiferromagnetic ordering on a bipartite lattice. 
Hence, the sublattice magnetization $M$ in the local axes of the 
spins in which all spins in the model state $| \Phi \rangle$ point in the negative $z$-direction 
is given by 
\begin{equation}
M = - \frac 1N \sum_{i=1}^N \langle \tilde \Psi \mid s_i^z \mid \Psi \rangle \; .
\label{eq10}
\end{equation}
This quantity is easily determined once the bra- and ket-state equations
have been solved. 

\subsection{Construction of CCM Valence-Bond Ket Ground States}
\label{sec:22}

Following on from our discussion above of the use of independent-spin product states as CCM model states, we now present a method for creating VBC states within the CCM from such independent-spin product model states. As an example, we consider the Heisenberg model that has a Hamiltonian defined by 
\begin{equation}
H = \sum_{\langle i,j\rangle}  {\bf s}_i ~ \cdot ~  {\bf s}_j  ~~ ,
\label{heisenberg}
\end{equation}
where the indices $i$ and $j$ run over all lattice sites on a given lattice. The brackets around $\langle i,j\rangle$ indicate that all  
nearest-neighbor pairs are counted once and once only. For the bipartite lattices
considered here (namely, the linear chain, the square lattice, 
and the ``CAVO'' lattice), we choose a model state in which nearest-neighbor spins are antiparallel along, say, the $z$-direction. The local frames of the ``up'' spins are rotated by 180$^{\circ}$ so that they point downwards in these local axes. This is achieved by carrying out the following transformation of the local axes of these spins: $s^x \rightarrow -s^x$; $s^y \rightarrow s^y$; and, $s^z \rightarrow -s^z$. The model state is now formed from a product purely 
of ``down" spins in the  rotated spin 
coordinates, as described above. 

The Heisenberg Hamiltonian is now written in terms of the new spin axes by,
 \begin{equation}
H = - \sum_{\langle i,j\rangle} \bigl ( s_i^z s_j^z 
+ \frac 12 \{  s_i^- s_j^- + s_i^+ s_j^+ \} \bigr ) ~~ .
\label{heisenberg2}
\end{equation}
%The CCM creation operators are now written in terms of sums of products of single 
%spin-raising operators, $s^+_k \equiv s^x_k + {\rm i} s^y_k$, 
%(again with respect to their local spin axes), such that  
%%%%%%%%%%%%%%%% 
%\begin{equation}
%C_I = s^+_{i_1} s^+_{i_2} \cdots s^+_{i_l}  ~~ .
%\label{eq12}
%\end{equation}
%%%%%%%%%%%%%%%% 
%The coefficients ${\cal S}_I$, for example, now represent the spin-correlation coefficients specified by the sets of site indices, $\{i_1\}$, $\{i_1, i_2\}$ etc., on the regular lattices under consideration. 
We note that the total spin in the ``global'' $z$-direction, $s_T^z = \sum_i^N s_i^z$, 
is a conserved quantity of the ground state in all of the models studied here.

%As mentioned above,  valence-bond model states have been 
%previously used by creating a whole new matrix formalism for each new 
%case. 
%This has typically involved the direct use of, e.g., products of 
%dimer-singlet states 
%between nearest-neighbor 
%spins.
%The interested reader 
%is referred to Ref.  \cite{ccm6} for more details of this approach. 
%However, we note that these calculations are much more complicated and 
%%time-consuming to implement than 
%those for the  independent-spin product model states.
%Fortunately, there is a much simpler approach that also uses  
%independent-spin product model states, but
%constructs special dimerized (or quadrumerized) solutions of the 
%CCM equations.
%This approach is enabled by the recent advances in the basic CCM 
%implementation that allows 
%spin models for {\it any} crystallographic lattice to be studied.

We now present the method for creating VBC states from 
 independent-spin product model states.
%For example,
Let us consider for a moment the one-dimensional spin-half 
$J_1$--$J_2$ Heisenberg
antiferromagnet.  The relevant model state is the collinear 
N\'eel state \cite{mg5,ccm7,ccm29,ccm30}. 
However, as mentioned above, the ground state for 
$J_2/J_1$ near to 0.5  breaks the translational lattice symmetry 
and is two-fold degenerate
(see also Sect.~\ref{results_1d}). 
In order to take into account this
property, we have to double the unit cell, i.e.    
the relevant unit cell has two neighboring sites at points (0,0,0) and (1,0,0) and a 
single Bravais vector of
(2,0,0)$^T$. Note, however, that such an explicit increase of the unit 
cells is not
necessary for VBC phases that do not break the
translational lattice symmetry.  
 The doubling of the unit cell, now enables us 
to consider two distinct types of two-spin nearest-neighbor 
ket-state correlation coefficients; in this way allowing 
to break the translational lattice symmetry. We 
call the two nearest-neighbor 
ket-state correlation coefficients 
${\cal S}_2^a$ and ${\cal S}_2^b$, where, 
we define ${\cal S}_2^a$ to connect those nearest-neighbor sites 
between different unit cells and ${\cal S}_2^b$ to connect 
those nearest-neighbor sites within each unit cell. 
With respect to the rotated spin coordinates, 
we may now construct   via Eq. (\ref{eq2}) a
simple dimerized product CCM ket state 
given 
by either ${\cal S}_2^a=1$ and ${\cal S}_2^b=0$ or  
${\cal S}_2^a=0$ and ${\cal S}_2^b=1$ for 
$S_2 = {\cal S}_2^a \sum_{i_a} s_{i_a}^+ s_{i_a+1}^+ 
+ {\cal S}_2^b \sum_{i_b} s_{i_b}^+ s_{i_b+1}^+$, and where $i_a$ runs 
over all sites with odd-numbered indices and $i_b$ runs over all 
sites with even-numbered indices. 
It is obvious, that this choice: (i) breaks the lattice
symmetry; and, (ii) represents two different degenerate states.
The proof that the above choice for ${\cal S}_2^a$, ${\cal S}_2^b$ 
leads to dimerized product states is as follows:
\begin{eqnarray} 
|\Psi\rangle &=& {\rm e}^{S_2} |\Phi\rangle ~~ \nonumber \\
                     &=& {\rm e}^{({\cal S}_2^a s_1^+ s_2^+ +  {\cal S}_2^b s_2^+ s_3^+ + {\cal S}_2^a s_3^+ s_4^+ + {\cal S}_2^b s_4^+ s_5^+ \cdots )} |\Phi\rangle ~~ \nonumber \\
{\cal S}_2^a=1~;~{\cal S}_2^b=0 ~ \Rightarrow ~  |\Psi\rangle                 &=& {\rm e}^{(s_1^+ s_2^+  + s_3^+ s_4^+ +  s_5^+ s_6^+ + \cdots )}  |\Phi\rangle ~~ \nonumber \\
                     &=& (1+ s_1^+ s_2^+)  (1+s_3^+ s_4^+)  (1+ s_5^+ s_6^+) \cdots  |\Phi\rangle ~~ \nonumber \\
                     &=& \{ |\downarrow_1 \downarrow_2 \rangle + | \uparrow_1 \uparrow_2 \rangle \} \otimes
                              \{ |\downarrow_3 \downarrow_4 \rangle + | \uparrow_3 \uparrow_4 \rangle \}  \nonumber \\
                       &&        \otimes 
                            \cdots 
                               ~~ 
                              \label{dimerstate}
\end{eqnarray}
where the notation $|\downarrow_i  \rangle$ and $|\uparrow_i  \rangle$ 
indicates a `down' or `up' spin, respectively, localized to site $i$. 
We note that $(s_i^+)^2|\Phi\rangle=0 ~ \forall ~ i$ is assumed in 
Eq.~(\ref{dimerstate}), 
which holds true for spin-half systems such as those considered here.

Furthermore, we note that if the local axes of spins on one sublattice 
are ``re-rotated'' such 
that `down' spins become `up' spins once again, i.e., so as to regain the 
Heisenberg Hamiltonian of Eq. (\ref{heisenberg}) in the global spin coordinate system, 
then the  dimerized product state becomes the usual 
product of the nearest-neighbor dimer singlets, 
namely, ($|\uparrow_i \downarrow_j \rangle - |\downarrow_i \uparrow_j \rangle$) 
at nearest-neighboring sites $i$ and $j$. 
Note that if such a dimerized product state becomes 
the true ground state 
of a given spin problem 
(e.g., the spin-half linear chain $J_1$--$J_2$ model at $J_2/J_1$=0.5 -- 
see below for more details) 
 then for any level of LSUB$m$ ($m \ge 2$) 
approximation either ${\cal S}_2^a=1$ with {\it all other} 
ket-state correlation coefficients (i.e., including those for $m$ spins with $m \ge 2$) 
equal to zero or ${\cal S}_2^b=1$ with {\it all other} ket-state correlation coefficients 
equal to zero, is a valid ket ground-state solution to this problem. 

The situation  is different for the bra state. Firstly, we note that 
 the dimerized product bra state being equivalent to the corresponding 
dimerized  ket product state of Eq. (\ref{dimerstate}) 
in rotated coordinates ought to be,
\begin{eqnarray} 
\langle \Psi | &=& \{ \langle \downarrow_1 \downarrow_2 | + \langle  \uparrow_1 \uparrow_2 | \} \otimes
                              \{ \langle \downarrow_3 \downarrow_4 | + \langle  \uparrow_3 \uparrow_4 | \} \otimes 
                            \cdots
                               ~~ \nonumber \\ 
                          &=&    \{ \langle \downarrow_1 \downarrow_2  \downarrow_3 \downarrow_4 
                                        	        \downarrow_5 \downarrow_6 \cdots | + \langle \uparrow_1 \uparrow_2 
	                                              \downarrow_3 \downarrow_4  \downarrow_5 \downarrow_6 \cdots |  +
			\langle \downarrow_1 \downarrow_2 \uparrow_3 \uparrow_4  \downarrow_5 \downarrow_6  \cdots |  + \nonumber \\
		       & \cdots &    + \langle \downarrow_1 \downarrow_2  \uparrow_3 \uparrow_4 
                                        	        \uparrow_5 \uparrow_6 \uparrow_7 \uparrow_8 \cdots | + \cdots  + 
	                                              \langle \uparrow_1 \uparrow_2  \uparrow_3 \uparrow_4  \uparrow_5 
	                                              \uparrow_6 \uparrow_7 \uparrow_8 \cdots | 
                             \}.
                               ~~ 
                               \label{exactbradimerstate}
\end{eqnarray}
We notice now that the modes of action of the spin operators (leftwards) on the bra spin states for $s=1/2$ are,
\begin{eqnarray} 
\langle \downarrow | s^+ = 0  & ~~ ; ~~ &  \langle \downarrow | s^- = \langle \uparrow | \nonumber \\
\langle \uparrow | s^- = 0  & ~~ ; ~~ &  \langle \uparrow | s^+ = \langle \downarrow |
.
\end{eqnarray} 
Thus, the NCCM ground bra state for LSUB$m$ with $m$ a finite number can only ever contain a maximum of $m$ ``up'' states 
in the bra state because of the linear 
nature of the bra-state operator $\tilde S$ in Eq. (\ref{eq2}).  
Hence, by contrast to the ket state, within the LSUB$m$ approximation 
we 
can never construct an equivalent simple dimerized product 
 bra state 
using the NCCM except in the exact limit
where $m \rightarrow \infty$. However, carrying out CCM calculations 
in the limit $m \rightarrow \infty$ using computational methods 
is generally not practical. We note that this problem might be 
alleviated by using the extended coupled cluster method (ECCM). In this method, 
the bra state 
is written in terms of an exponential with respect to both the ket-state {\it and} 
bra-state correlation operators. It is this exponential term that allows such dimer solutions, 
e.g., for the ket state for the NCCM above. However, we do not discuss the ECCM further
in this present paper.

 We mention that  the sublattice magnetization $M$
(see Eq. (\ref{eq10})) for the simple dimerized product ket state  
(i.e.,  ${\cal S}_2^a=1$ and all other coefficients equal to zero) is written in terms of only 
one bra-state correlation coefficient, 
i.e. $M=\frac{1}{2} - 2\tilde {\cal S}_2^a$.

%Finally, we note again that the CCM bra and ket states are not necessarily Hermitian conjugates at 
%any level of approximation, except for the case where $m \rightarrow \infty$. 
%Indeed, this behavior is what allows the possibility that the ``approximated'' ket state might be an exact 
%ground state and the corresponding approximated 
%bra state might not. However, the behavior of the bra state is strongly 
%case-dependent and it is investigated in more detail for some specific lattice spin models 
%below. 

\begin{figure}
\epsfxsize=11cm
\centerline{\epsffile{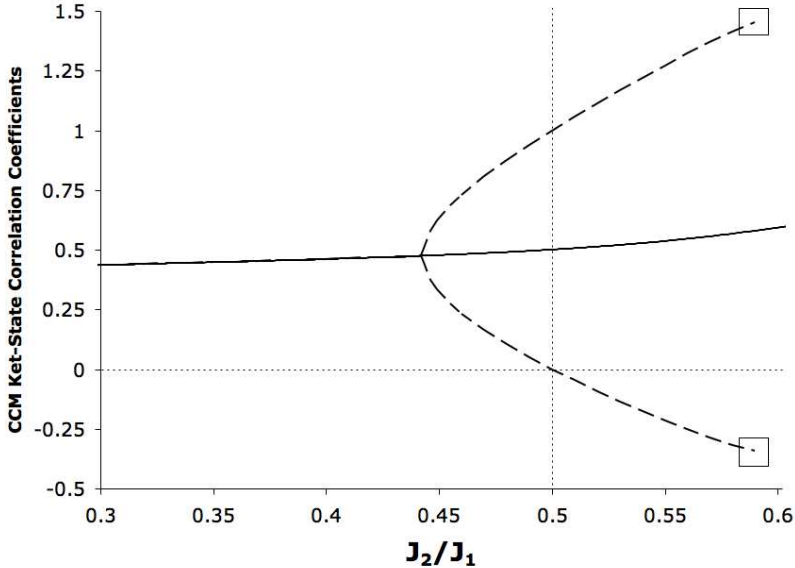}}
\caption{CCM results at the LSUB12 level of approximation for 
the ground-state nearest-neighbor ket-state 
correlation coefficients of the spin-half $J_1$--$J_2$  
antiferromagnet on the linear chain. 
The nearest-neighbor coefficients ${\cal S}_2^a$ and  ${\cal S}_2^b$ of the
 symmetry breaking dimerized solution are
shown by dashed lines.
Results for the usual (`N\'eel-type') solution
(where ${\cal
S}_2^a={\cal S}_2^b$) are shown by the full line. 
Below the (bifurcation) CCM critical point at $J_2/J_1|_{c_1}$
there is only the solution with ${\cal
S}_2^a={\cal S}_2^b$. 
A termination point $J_2/J_1|_{t}$ of the CCM equations 
for the dimerized solution, 
at which point the real solution to the CCM equations
terminates, is indicated by the boxes. }
\label{j1_j2_ket_coeffs}
\end{figure}

\section{Results}
\label{sec:3}

 \subsection{The Spin-Half $J_1$--$J_2$ Model on the Linear
Chain}\label{results_1d}

The Hamiltonian for this spin-half model has nearest-neighbor bonds of 
strength $J_1$ and next-nearest-neighbor bonds of strength $J_2$. We use 
a N\'eel model state in which nearest-neighbor spins on the linear chain are antiparallel. We rotate the spin coordinates of the `up' spins so that notationally they become `down' spins in these locally defined axes. The relevant Hamiltonian in rotated coordinates is then given by
\begin{equation}
H = -J_1 \sum_{\langle i,j\rangle} \bigl ( s_i^z s_j^z + \frac 12 \{  s_i^- s_j^- + s_i^+ s_j^+ \} \bigr )  
       + 
        J_2 \sum_{\langle \langle i,k \rangle \rangle} \bigl ( s_i^z s_k^z + \frac 12 \{  s_i^+ s_k^- + s_i^- s_k^+ \} \bigr ) ~~ ,
\label{j1j2hamiltonian}
\end{equation}
where $\langle i,j\rangle$ runs over all nearest-neighbor sites on the lattice counting each pair once and once only and $\langle \langle i,k \rangle \rangle$ runs 
over all next-nearest-neighbor sites on the lattice, again counting each pair once and once only. Henceforth we put $J_1=1$ and consider $J_2>0$. 

\begin{figure}
\epsfxsize=11cm
\centerline{\epsffile{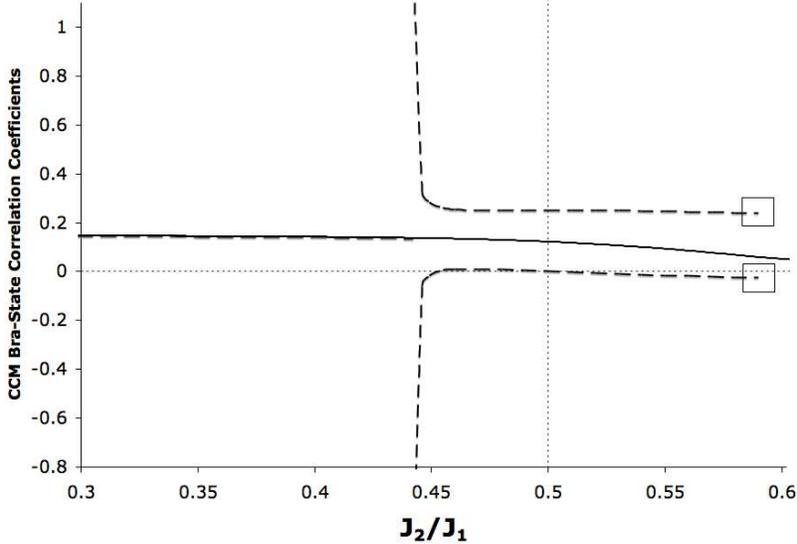}}
\caption{CCM results at the LSUB12 level of approximation for the ground-state nearest-neighbor bra-state 
correlation coefficients of the spin-half $J_1$--$J_2$  antiferromagnet on the linear chain. The nearest-neighbor coefficients 
$\tilde {\cal S}_2^a$ and  $\tilde {\cal S}_2^b$ of the dimerized solution are shown by dashed lines.
Results for the  usual (`N\'eel-type') solution 
(where $\tilde {\cal S}_2^a= \tilde {\cal S}_2^b$) are shown by the full line.
Below the critical point at $J_2/J_1|_{c_1}$ both solutions coincide. 
Results for these bra-state correlation coefficients diverge at  the
critical point $J_2/J_1|_{c_1}$. 
A termination point at $J_2/J_1|_{t}$ is shown by the boxes on the right-hand side of the figure. }
\label{j1_j2_bra_coeffs}
\end{figure}

The ground-state properties of this system have been studied using methods such as 
exact diagonalizations  \cite{mg2,aligia00}, DMRG  \cite{mg3,mg4,mg5,ccm7}, CCM
 \cite{ccm4,ccm6,ccm29}, and field-theoretical approaches   \cite{mg5} (see Refs. 
 \cite{mg5,mg6} for a general review).  Note that
previous CCM studies of the model 
 considering only independent-spin product model states that conserve 
the lattice symmetry
are reported in
Refs.~ \cite{ccm4,ccm30}.
At $J_2/J_1=0$ we have the 
unfrustrated Heisenberg antiferromagnet, where the exact solution is provided by the Bethe Ansatz.
The ground state is gapless and the spin-spin 
correlation function $\langle  {\bf s}_i  \cdot {\bf s}_j\rangle$  decays slowly to zero according to a power-law, 
i.e. no true N\'eel-like long-range order is observed. 
In the region $J_2/J_1>0$ the nearest-neighbor ($J_1$) and next-nearest-neighbor
interactions ($J_2$)
compete, thus leading to frustration. At $J_2/J_1 = 0.2411(1)$
the model exhibits a transition 
to a two-fold degenerate gapped dimerized phase with an exponential decay of the
correlation function $\langle  {\bf s}_i \cdot 
{\bf s}_j\rangle$ \cite{mg2,mg3,mg5,mg6}. This state
breaks the translational lattice symmetry. At the Majumdar-Ghosh 
point $J_2/J_1=0.5$ 
there are two degenerate   simple exact dimer-singlet product ground 
states corresponding to the dimerized product state 
of Eq.~(\ref{dimerstate}) for the Hamiltonian of Eq. (\ref{j1j2hamiltonian}) \cite{mg1}. 
(We recall that rotated spin coordinates are used in Eq.~(\ref{dimerstate}).)

\begin{figure}
\epsfxsize=11cm
\centerline{\epsffile{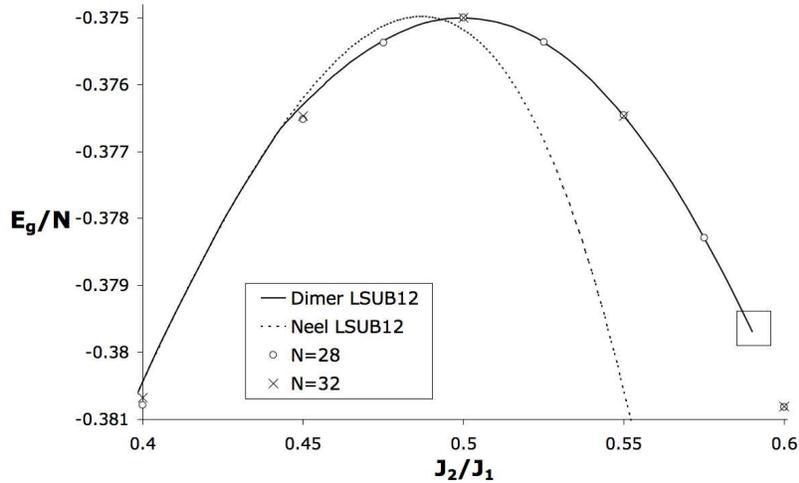}}
\caption{CCM results for the ground-state energy of the spin-half $J_1$--$J_2$  antiferromagnet 
with $J_1=1$ on the linear chain at the LSUB12 level of approximation. 
The dimerized and usual (`N\'eel-type') solutions are shown in this figure. Results of exact diagonalizations for $N=28$ and $N=32$ number of sites is also shown. 
The CCM termination point for the dimerized solution is shown by the box.}
\label{j1_j2_energies}
\end{figure}

We now consider how this model can be treated in the dimerized phase by the CCM 
via, as presented above, the identification of a  special dimerized 
solution of 
the CCM equations 
for a N\'eel model state. Also as discussed above,
we use a doubled unit cell including two neighboring sites for a spin-half 
system on the linear chain at points (0,0,0) and (1,0,0) and 
a single Bravais vector 
(2,0,0)$^T$ to take into account the symmetry breaking. 
There are thus two distinct types of two-spin 
nearest-neighbor ket-state correlation coefficients and again 
these are denoted as ${\cal S}_2^a$  and ${\cal S}_2^b$.  
The exact ground state at $J_2/J_1=0.5$ of Eq. (\ref{dimerstate}) is obtained by setting 
${\cal S}_2^a=1$ and all other coefficients equal to zero. 
 Starting from $J_2/J_1=0.5$ we 
are able to track 
this exact
solution at $J_2/J_1=0.5$ within a certain LSUB$m$ approximation for other values of $J_2/J_1$ and 
the results for the 
nearest-neighbor ket-state correlation coefficients in LSUB12
approximation are presented in Fig. \ref{j1_j2_ket_coeffs}. Clearly,  we see that the 
exact dimerized product-state solution
for the ket ground state is obtained within LSUB12 level of approximation (and, indeed, 
at all LSUB$m$ approximation with $m \ge 2$)
at $J_2/J_1=0.5$, i.e. ${\cal S}_2^a=1$ and all other coefficients equal to zero. Moving
away from $J_2/J_1=0.5$ we still find a CCM ground state that breaks the 
lattice symmetry.
However, this dimerized state deviates from the simple product state,
i.e. ${\cal S}_2^a \ne 1$ and other non-zero coefficients ${\cal S}_I$ occur. 
\begin{figure}
\epsfxsize=11cm
\centerline{\epsffile{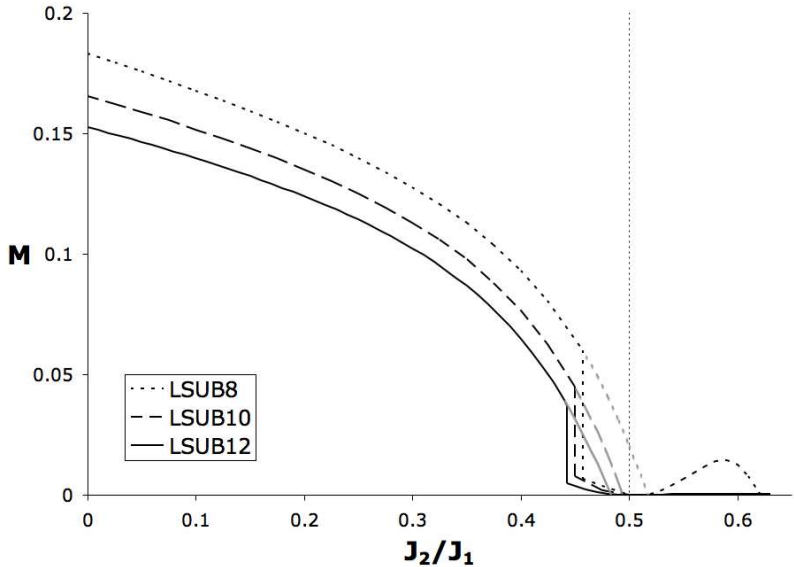}}
\caption{CCM results for the sublattice magnetization $M$ of the spin-half $J_1$--$J_2$  antiferromagnet on the linear chain.
Below the critical point at $J_2/J_1|_{c_1}$ the results for 
the  usual (`N\'eel-type')
and the dimerized  solution coincide. At $J_2/J_1|_{c_1}$ the sublattice
magnetization of the dimerized solution exhibits a jump,  whereas $M$ for the
usual (`N\'eel-type') solution is continuous. 
Above $J_2/J_1|_{c_1}$ the sublattice magnetization of the dimerized  solution
(black lines)
is smaller than that of the usual (`N\'eel-type') 
solution (grey lines), except for the LSUB8 in a small
region above $J_2/J_1 =0.5$.  
}
\label{j1_j2_mag}
\end{figure}
Furthermore,  the  solution (i.e. ${\cal S}_2^a={\cal
S}_2^b$)
having full translational symmetry  is the only solution below a critical
point 
$J_2/{J_1}|_{c_1}$($<0.5$). Henceforth, we shall refer to this solution below
$J_2/{J_1}|_{c_1}$ as the ``usual (`N\'eel-type') solution'' because previous
CCM calculations  \cite{ccm4,ccm6} for the $J_1$--$J_2$ model have 
considered the non-symmetry breaking case only. 
For larger values of $J_2/{J_1}$ a CCM termination point is observed at $J_2/{J_1}|_{t}$ ($>0.5$), 
shown by the boxes in Fig. \ref{j1_j2_ket_coeffs}. At this point, the real 
solution of the CCM dimerized solution is terminated. 
These CCM results indicate that a dimerized phase exists over a finite range of
$J_2/J_1$, which is in agreement with known results, see e.g.
Refs. \cite{mg3,mg5,mg6}. Qualitatively similar results are observed at other levels of LSUB$m$ approximation for the ket-state correlation coefficients as a function of $J_2/J_1$. The results for $J_2/{J_1}|_{c_1}$ and $J_2/{J_1}|_{t}$ are shown in Table \ref{tab1}. It is obvious that the CCM
critical point $J_2/{J_1}|_{c_1}$ becomes smaller (i.e., becomes closer to
the true critical point $J_2/J_1 = 0.2411(1)$ \cite{mg3,mg6})  with higher orders
$m$
of LSUB$m$ approximation. However,  the critical point $J_2/{J_1}|_{c_1}$ is still 
significantly too high  even at the LSUB12 level of approximation. 
However, as shown in
Ref.~ \cite{mg5}  
the dimerization gap for $J_2/J_1< 0.4$ is very small and it is
therefore not very surprising that we do not detect the dimerized phase
below $J_2/J_1< 0.4$ using LSUB$m$ approximations with $m \le 12$. On the side of
$J_2/J_1 > 0.5$ the existence of termination points can be related to the
appearance
of incommensurate spiral spin correlations at $J_2/J_1 >
0.6$ \cite{mg5,aligia00,ccm7,ccm29} that are
not taken into account in the model state used here.

The nearest-neighbor bra-state correlation coefficient at the LSUB$m$ level of approximation 
at $J_2/J_1=0.5$ has $\tilde {\cal S}_2^a=1/4$ with $m \ge 4$. This is shown in Fig. \ref{j1_j2_bra_coeffs} for the 
LSUB12 level of approximation. We find that the bra-state solution for the 
nearest-neighbor correlation coefficients is close to $1/4$ over the range $J_2/{J_1}|_{c_1} < J_2/{J_1} \le
J_2/{J_1}|_{t}$.  However,  we find that the 
nearest-neighbor correlation coefficient diverges as $J_2/{J_1} \rightarrow J_2/{J_1}|_{c_1}$ 
and this is also shown in Fig. \ref{j1_j2_bra_coeffs}. 
Again, the  usual (`N\'eel-type') 
solution  ($\tilde {\cal S}_2^a= \tilde {\cal
S}_2^b$)  is obtained for $J_2/{J_1}<J_2/{J_1}|_{c_1}$. 
The upper CCM termination point at
$J_2/{J_1}|_{t}$ is also shown in Fig. \ref{j1_j2_bra_coeffs} 
by the boxes on the right-hand side of the figure. 
Once more, qualitatively similar results are observed at other 
levels of LSUB$m$ approximation for the bra-state correlation coefficients 
as a function of $J_2/J_1$.

\begin{table}[t]
\caption{CCM results for the positions of the range of the dimerized phase. }
\begin{center}
{\footnotesize
\begin{tabular}{|c|c|c|}                                                                                     
\hline\noalign{\smallskip}
LSUB$m$		&$J_2/{J_1}|_{c_1}$   &$J_2/{J_1}|_{t}$                 \\ 
\noalign{\smallskip}\hline\noalign{\smallskip}
2			&0.4761                         &--        				          \\ 
4			&0.4745                         &0.5576            			 \\ 
6			&0.4637                         &--             				 \\ 
8			&0.4568                         &0.7410         				 \\ 
10			&0.4498                         &0.6404          			 \\ 
12			&0.4429                         &0.5956            			\\ 
\noalign{\smallskip}\hline
\end{tabular}
}
\end{center}
\label{tab1}
\end{table}

We now consider the ground-state energy of this system in the dimerized regime. Our results for 
the new dimer solution and the usual  (`N\'eel-type') solution 
are shown in Fig.~\ref{j1_j2_energies}. Firstly, we note that the exact ground-state energy of $E_g/N=-0.375 J_1$ is obtained at the point $J_2/J_1=0.5$, as expected. 
We note again that our solution is an exact ground eigenstate at this point.  
We see that ground-state energy of the 
usual  (`N\'eel-type') CCM solution  
in which ${\cal S}_2^a={\cal S}_2^b$ at the LSUB12 
level of approximation actually lies below this exact solution. 
This indicates (i) that the usual  (`N\'eel-type') CCM solution is 
a relatively poor choice at this
point; and, (ii)  that the CCM ground-state energy does not 
fulfill the variational
principle \cite{ccm13a}.
Furthermore,   we see that CCM dimer solution 
compares extremely well to results of exact diagonalizations for $N=28$ and $N=32$ sites in the dimerized regime 
shown in Fig.  \ref{j1_j2_energies}. It certainly provides far better results than those of 
the 
 usual (`N\'eel-type')
CCM solution 
beyond the critical point  at $J_2/{J_1}|_{c_1}$.
%The critical point  at $J_2/{J_1}|_{c_1}$ is shown by the ``kink" in the ground-state energies at this point. 
%The second derivative of the ground-state energy diverges at this point. Again, we note that there is also an upper critical point at
%$J_2/{J_1}|_{t}$.

The results for the sublattice magnetization $M$ of this model are presented graphically in Fig. \ref{j1_j2_mag}. 
Since the one-dimensional $J_1$--$J_2$ model does not possess N\'eel long-range
order for any value of $J_1,J_2 \ge 0$ the true value is $M=0$.
As is known from previous CCM
calculations \cite{ccm4,ccm15,ccm29,ccm30},  
the sublattice magnetization is nonzero (but small) using the  
usual N\'eel model state.
However,  the correct result $M=0$  can be obtained \cite{ccm29,ccm30} by
extrapolating the `raw' LSUB$m$ data to $m\to\infty$.
Indeed it is obvious that the CCM-LSUB$m$ values for $M$ are non-negligible for
N\'eel model state in the region $J_2/J_1 < J_2/{J_1}|_{c_1}$. It is also obvious that
$M$ decreases with the level of approximation $m$ approaching the true value
$M=0$, and, that increasing the strength $J_2$ of the frustration weakens magnetic order.
More interestingly, we find that the sublattice magnetization behaves 
discontinuously at $J_2/{J_1}|_{c_1}$ by 
tracking the lattice symmetry-breaking  dimerized 
solution, 
and then remains near to zero at LSUB10 and LSUB12 levels of approximation 
across the entire range $J_2/{J_1}|_{c_1}<J_2/{J_1}<J_2/{J_1}|_{t}$.  (At the 
lower LSUB8 level of approximation the results for the sublattice magnetization 
differ from zero by a small amount in a small region above
$J_2/{J_1}>0.5$.) On the other hand, by tracking the  usual (`N\'eel-type')
solution $M$ changes continuously with $J_2$ and it is
larger than for the dimerized solution for
$ J_2/{J_1}|_{c_1}<J_2/{J_1} < 0.5$.
This behavior of $M$ is another indication that the dimerized CCM solution
describes
the true physics of the model much better than the usual N\'eel solution.      
We note finally that the CCM sublattice magnetization is exactly zero at 
the Majumdar-Ghosh point $J_2/J_1=0.5$ at all levels of LSUB$m$ approximation using the dimerized product state.

We remark again that our results for the lower phase transition 
point $J_2/{J_1}|_{c_1}$ overestimate the position by a factor of two.
We re-iterate that we believe that we over-estimate $J_2/{J_1}|_{c_1}$ 
because the energy gap only becomes large for values of $J_2/J_1$ 
of approximately 0.4 \cite{mg5}. Finally, we note that results given
here present the possibility that the CCM might be applied to detect 
spontaneous symmetry breaking for systems of two or three spatial
dimensions, i.e., where other approximate methods become less accurate or 
may not even be applicable (e.g., such as the DMRG method). 

\subsection{The Shastry-Sutherland Antiferromagnet}
\label{results_shastry}

\begin{figure}
\epsfxsize=7cm
\centerline{\epsffile{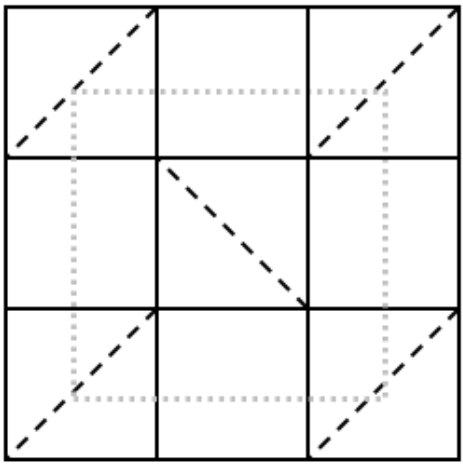}}
\caption{The nearest-neighbor bonds (solid lines) of strength $J_1$ and the next-nearest-neighbor
diagonal bonds (dashed lines) of strength $J_2$ for the Shastry-Sutherland model. The geometric unit cell is shown by the square with the grey dotted lines.}
\label{shastry_lattice_bonds}
\end{figure}

Another model that demonstrates dimer order is the Shastry-Sutherland antiferromagnet 
 \cite{shastry}.  The Shastry-Sutherland antiferromagnet is a spin-half 
Heisenberg model on an underlying square lattice with antiferromagnetic nearest-neighbor bonds $J_1$ and with 
one antiferromagnetic next-nearest-neighbor diagonal  bond $J_2$ in every second square
(plaquette), as shown in Fig. \ref{shastry_lattice_bonds}. We note that for bonds of equal strength, 
i.e., $J_1=J_2$,  the Shastry-Sutherland model is equivalent to a Heisenberg 
model  
on one of the eleven uniform Archimedean lattices  \cite{shastry2}. Interest in this 
model has been 
renewed by the discovery of the magnetic material $\mathrm{SrCu}(\mathrm{BO}_3)_2$  \cite{shastry2,shastry3} that can be understood 
in terms of the Shastry-Sutherland model. The ground state of this model in the limit of small frustration
$J_2/J_1 \ll 1$ and 
large frustration $J_2/J_1 \gg 1$ is well understood. However, the ground-state phase at 
intermediate values of $J_2/J_1 \approx 1$  is still a matter of discussion.

Just as in the case of the one-dimensional $J_1$--$J_2$ model
this model also has a simple exact  dimer-singlet product
ground state
in a certain parameter region. 
However, by contrast with the one-dimensional $J_1$--$J_2$ model
the  dimer-singlet product
ground state of the Shastry-Sutherland model is built up of a product 
of dimer singlets 
located on the next-nearest $J_2$ bonds and does not break the translational 
symmetry. 

This  model has been treated  previously by Schwinger boson mean-field theory  \cite{shastry3}, exact diagonalization of small lattices 
 \cite{shastry2,shastry4,shastry5}, series expansions  \cite{shastry6,shastry7,shastry8,shastry9}, 
the renormalization group \cite{shastry9}, a gauge-theoretical approach  \cite{shastry10}, 
and the CCM  \cite{shastry15,ccm29}. 
A recent review can be found in Ref.  \cite{shastry11}.  We know from these studies that the physics of 
the quantum model is  similar to that of its classical counterpart  for small $J_2
< J_1$, i.e., we have 
semi-classical N\'eel long-range order. Furthermore, we know  
\cite{shastry} that a
 simple dimer-singlet
product state 
given by dimer singlets on the diagonal bonds indicated by the dashed 
lines in Fig. \ref{shastry_lattice_bonds} is the quantum ground state for 
large  $J_2$. The energy per site of this  dimer-singlet
product state state is $E_{{\rm dimer}}/N= -3J_2/8$. It  becomes the
ground state for $J_2 > J_2^c$ where $J_2^c \approx (1.465 \pm 0.025) J_1$. Note that, by
contrast to the one-dimensional $J_1$--$J_2$ model, the transition to the
dimerized phase in the Shastry-Sutherland model is most likely of first
order.   

The application of the CCM to this model has been discussed at length in 
Ref.~\cite{shastry15}. The interested reader is referred 
to this reference for more details about both the model and the details 
of the applying the CCM to it. 
We note that the  CCM solution for the  
N\'eel model state with nearest-neighbor spin antiparallel was identified and this 
worked well in the region of $J_2/J_1 < 1.6$. This case has a 
Hamiltonian similar to that of Eq. 
(\ref{heisenberg2}), the nearest-neighbor bonds $J_1$ (solid lines) 
and next-nearest-neighbor bonds $J_2$ (dashed lines) 
run over those sites on the square lattice, as opposed to the linear 
chain for the case presented above.

However, we will show that a simple dimerized product ket 
state solution to the CCM equations also exists. 
Note firstly that we may define a  collinear 
independent-spin product model state. 
This is a model
state in which next-nearest-neighbor spins are antiparallel.
This state is relevant for large antiferromagnetic $J_2$,
since antiparallel next-nearest-neighbor spins satisfy the $J_2$ bonds.  
 Thus, the ``up'' and ``down'' spins form alternate neighboring 
columns (or rows), see e.g. 
Ref.~\cite{ccm31,ccm32,ccm33,ccm34,ccm35}.  
We choose the former case and call the corresponding model state the 
`columnar model state'.
We rotate the ``up'' spins into (nominally) ``down''  as illustrated in
Sect.~\ref{ccm_f}, although we must now also take the columnar 
form of the model state into account. 
The relevant Hamiltonian in the appropriate local axes described above is given by 
\begin{eqnarray}
H &=& -J_1 \sum_{\langle {i_x},j_x\rangle} \bigl ( s_{i_x}^z s_{j_x}^z + \frac 12 \{  s_{i_x}^- s_{j_x}^- + s_{i_x}^+ s_{j_x}^+ \} \bigr )  \nonumber \\
       && +J_1 \sum_{\langle {i_y},{j_y}\rangle} \bigl ( s_{i_y}^z s_{j_y}^z + \frac 12 \{  s_{i_y}^- s_{j_y}^+ + s_{i_y}^+ s_{j_y}^- \} \bigr )  
       \nonumber \\
       & &
            - J_2 \sum_{\langle \langle i,k \rangle \rangle} \bigl ( s_i^z s_k^z + \frac 12 \{  s_i^- s_k^- + s_i^- s_k^- \} \bigr ) ~~ ,
\label{shastryhamiltonian}
\end{eqnarray}
where the sum on $\langle {i_x},j_x\rangle$ runs over all nearest-neighbor 
pairs of the lattice sites  in the row or $x$-direction and the sum on 
$\langle {i_y},{j_y}\rangle$ runs over all nearest-neighbor pairs
of the lattice sites in the column or $y$-direction. Furthermore, the sum on $\langle \langle i,k \rangle \rangle$ runs over distinct next-nearest-neighbor pairs of sites connected by the broken lines in Fig. \ref{shastry_lattice_bonds}. We count each bond once and once only (for both the nearest-neighbor and next-nearest-neighbor bonds). The unit cell for this model contains four sites, and it is shown also in Fig. \ref{shastry_lattice_bonds}. Once again, in what follows we set $J_1=1$ and treat $J_2 > 0$ as the parameter of interest in the model.

\begin{figure}
\epsfxsize=11cm
\centerline{\epsffile{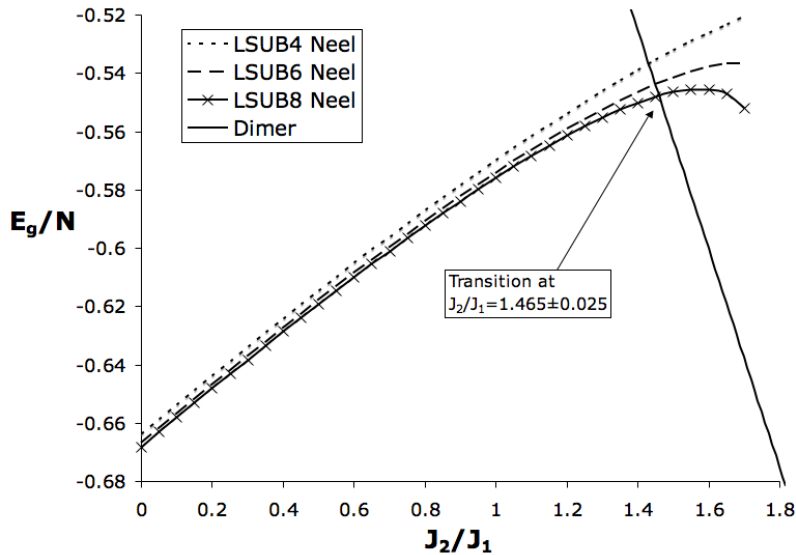}}
\caption{The CCM ground-state energy for the spin-half Shastry-Sutherland model 
(with $J_1=1$) using the N\'eel model state (LSUB4, LSUB6, LSUB8) and the 
columnar model state. Note that in the latter case the exact ground-state energy 
is obtained for any LSUB$m$ with $m\ge2$. }
\label{shastry_energies}
\end{figure}

At $J_2=0$, we have the usual square-lattice antiferromagnet, 
and the system is N\'eel-ordered at this point and in a finite range 
for $J_2>0$. 
Furthermore, we note that the ground state becomes the 
 simple dimer-singlet product state 
for $J_2 > J_2^c$ where $J_2^c \approx (1.465 \pm 0.025) J_1$. 
We are able to define CCM 
correlation coefficients for spin dimers located on the diagonal bonds of 
the dashed lines of Fig. \ref{shastry_lattice_bonds} 
with respect to the columnar model state. By setting these 
ket-state correlation coefficients to unity and all other CCM 
multispin correlation coefficients to zero, we are able to form the relevant
dimer-singlet product state that is the exact ground state in this regime. 
We find that this CCM dimer solution is a stable solution for 
the CCM equations for all values of $J_2/J_1>0$, and this is because
the  dimer-singlet product  state is a true eigenstate for any values of $J_1$ and
$J_2$. However, the energy of this state is low enough for it to 
become the ground-state energy only for large $J_2$.  The CCM dimer solution yields 
also the correct exact energy 
for the  dimer-singlet product  state, namely, $E_g/N=-0.375J_2$. 
The CCM ground-state energies are now shown for both the dimer and 
usual N\'eel results in Fig. \ref{shastry_energies}.  
We see that there is a 
crossing of the N\'eel and dimer energies
at $J_2/J_1 \approx 1.48$, as reported in 
Ref.  \cite{ccm22}.

\begin{figure}
\epsfxsize=11cm
\centerline{\epsffile{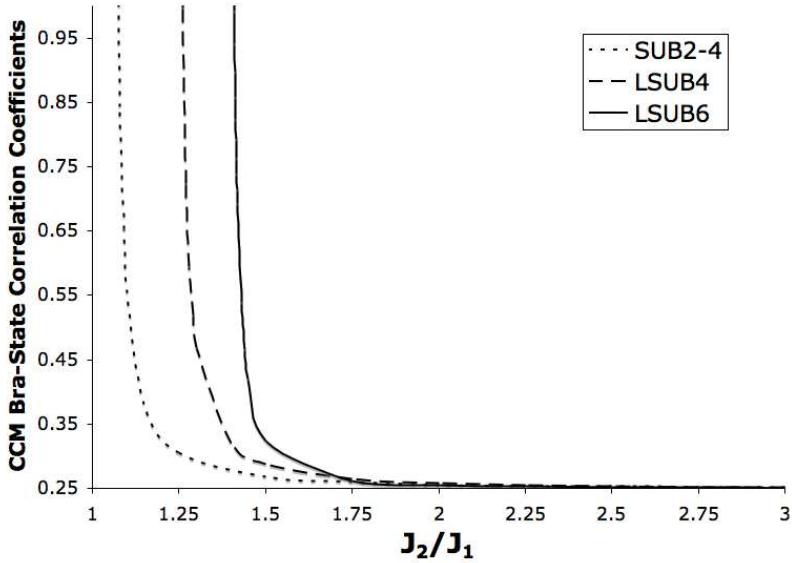}}
\caption{CCM results using the columnar model state
for the ground-state bra-state correlation coefficients 
for those sites connecting the dimers in the 
spin-half Shastry-Sutherland model.}
\label{shastry_bra_coeffs}
\end{figure}

Another interesting point is that the bra-state correlation coefficients 
do not remain constant with respect to varying $J_2$. Indeed, we find many of the bra-state correlation 
coefficients diverge at a CCM critical point as may be seen in Fig. \ref{shastry_bra_coeffs} for the next-nearest-neighbor bra-state correlation 
coefficients on the diagonals (i.e., those corresponding to the dashed
lines in Fig. \ref{shastry_lattice_bonds}). 
This is  a critical point for the CCM bra-state equations only. 
The ket-state equations clearly do not contain a similar critical point. 
Hence, this is a critical point that is ``driven'' by the bra state alone.  
The critical points occur at values for $J_2/J_1$ of $1.059$, $1.243$, and $1.397$ for the SUB2-4, LSUB4, 
and LSUB6 levels of approximation, respectively. This is in agreement with the position of the phase transition point at $J_2^c \approx (1.465 \pm 0.025) J_1$. We note that the CCM ket state is an 
exact ground 
eigenstate in this regime, whereas the bra state is not. As mentioned before, this is because the ket states and bra states are not explicitly 
constrained to be Hermitian conjugates of each other in the 
CCM parametrizations.

\begin{figure}
\epsfxsize=11cm
\centerline{\epsffile{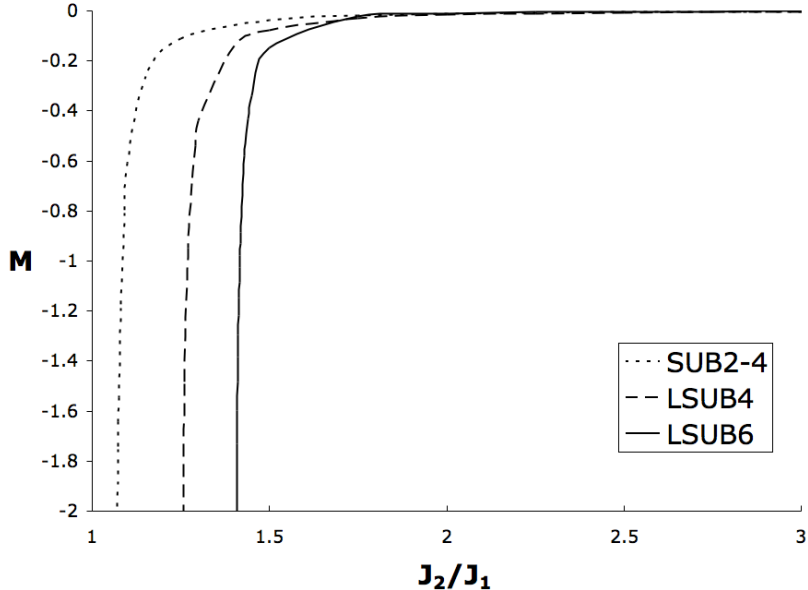}}
\caption{The sublattice magnetization for the spin-half Shastry-Sutherland 
model for the  columnar model state.}
\label{shastry_mag}
\end{figure}

The results for the sublattice magnetization $M$ (with respect to the columnar model state)  
are shown in Fig. \ref{shastry_mag}. 
We see that the values for the sublattice magnetization are negative for all values of $J_2/J_1$.
Note that the true values for the sublattice magnetization are actually zero in this regime. 
Negative values of $M$ might be an indication of missing magnetic long-range
order.  
However, we note that although the CCM results are negative, they remain close to zero 
(e.g., $|M| < 10^{-2}$ at the LSUB6 level of approximation) for $J_2/J_1>2$. 
This indicates that we obtain generally good results for the sublattice magnetization. 
However, we see  from Fig. \ref{shastry_mag} that our results clearly 
become worse for $J_2/J_1<2$. For example, we note that the sublattice magnetization diverges as 
we approach the CCM critical point (of the bra-state equations only) at $J_2^c/J_1$. 
We note that the bra and ket states are not explicitly constrained to 
be Hermitian conjugates. Hence, the bra state does not have to be an 
exact eigenstate of the Hamiltonian even though the ket state is 
for the Shastry-Sutherland model. We note therefore that this 
is enough to allow the bra-state equations to become critical 
even though the ket-state equations do not; hence, the nearest-
neighbour bra-state correlation coefficient ($\tilde {\cal S}_2^a$)
diverges. The sublattice magnetization ($M=\frac{1}{2} - 2\tilde 
{\cal S}_2^a$) therefore diverges also.
Again, this is a reflection of the critical point that is ``driven" by the 
bra state alone.
 The N\'eel state with nearest-neighbor spins antiparallel is the 
appropriate CCM model state \cite{shastry15} below the critical point $J_2<J_2^c$.
\subsection{The $J$--$J'$ 
Heisenberg antiferromagnet on the CAVO lattice}
\label{cavo_n_n}

\begin{figure}
\epsfxsize=7cm
\centerline{\epsffile{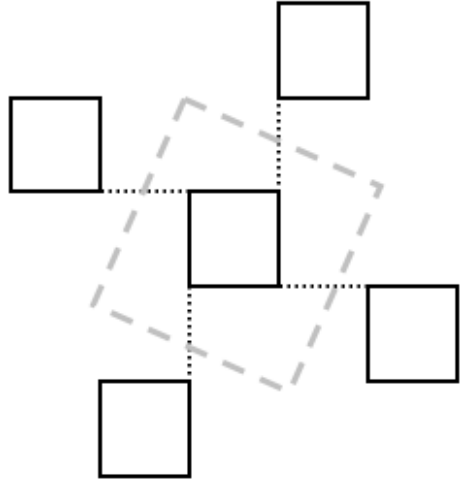}}
\caption{The CAVO lattice.
The nearest-neighbor bonds that connect two sites 
on a four-site plaquette are shown by the solid 
lines and have a bond strength 
given by $J$ (=1). The nearest-neighbor bonds that connect two sites 
on different plaquettes (dimer bonds) are shown by the dotted lines 
and have a bond strength 
given by $J'$. The unit cell of the lattice is shown by the square with the grey dashed lines.}
\label{cavo_unit_cell}
\end{figure}

 In this section, we consider an antiferromagnetic Heisenberg model 
in which the basic 
geometric unit 
cell 
contains four neighboring lattice sites on the underlying 
crystallographic lattice of the 
magnetic material CaV$_4$O$_9$ (CAVO), shown in Fig. \ref{cavo_unit_cell}.
There are two non-equivalent antiferromagnetic 
nearest-neighbor bonds $J$ and $J'$ 
belonging to dimers ($J'$) and to 
four-spin plaquettes ($J$) respectively.  The ground state of the quantum 
model 
depends on the ratio $J'/J$ of the competing bonds.  
Using a unit cell as defined in Fig.~\ref{cavo_unit_cell}, the plaquette bonds
$J$ are inside the four-site unit cell and the dimer bonds $J'$ connect sites in different unit
cells. We note that this model is not frustrated but the two  non-equivalent
nearest-neighbor bonds lead to a competition in the quantum system.  
Henceforth, we choose an energy scale such that $J=1$.

\begin{figure}
\epsfxsize=10cm
\centerline{\epsffile{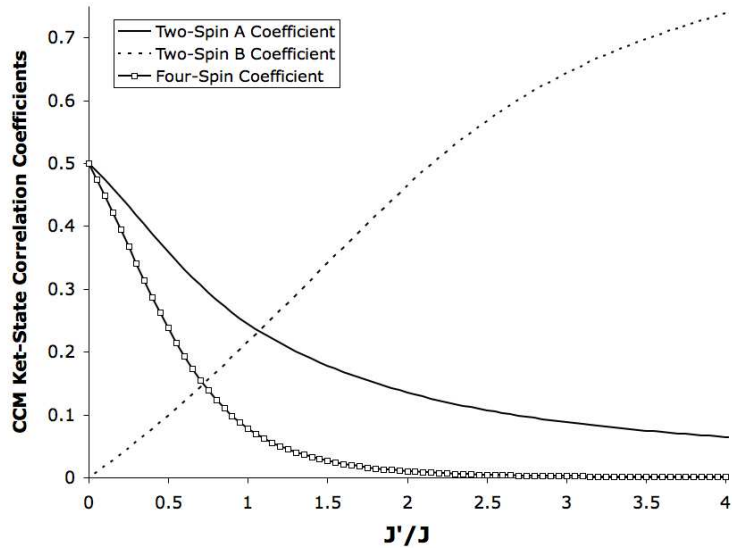}}
\caption{CCM results at the LSUB8 level of approximation for the ground-state nearest-neighbor
two-spin $A$ and $B$ ket-state coefficients (${\cal S}_2^a$ and  ${\cal S}_2^b$)
and the plaquette four-spin ket-state correlation (${\cal S}_4^p$) 
coefficients of the spin-half antiferromagnet on the
CAVO lattice.}
\label{cavo_ket_coeffs}
\end{figure}

We note that several techniques  \cite{cavo1,cavo2,cavo2a,cavo3,cavo4,cavo5,cavo6} suggest 
that the N\'eel-ordered ground state at $J'/J=1$ persists over a 
finite range of values of $J'/J$ around this point, $J'_{c_1}/J < J'/J < J'_{c_2}/J$. 
The best estimates for $J'_{c_1}/J$ and $J'_{c_2}/J$  are probably provided by quantum Monte Carlo 
calculations  \cite{cavo2,cavo2a} that suggest that the range over which N\'eel order is stable is given 
by $J'_{c_1}/J \approx 0.939$ and  $J'_{c_2}/J \approx 1.68 \pm 0.14$.  
For $J'/J<J'_{c_1}/J$ the model exhibits
a  quadrumerized plaquette VBC phase  with enhanced spin 
correlations on the four-spin plaquettes, 
and for $J'/J>J'_{c_2}/J$ it has
a dimerized VBC phase  with enhanced spin correlations on the dimers. 
Neither of the VBC
phases breaks the translational symmetry of the lattice.
Furthermore, we mention that
by contrast with the previously considered models (see 
Secs.~\ref{results_1d} and \ref{results_shastry})  
this model has no simple exact product ground state for any value of $J'/J$.
The interested reader can find more information on the ground-state phases 
in Refs.  \cite{cavo1,cavo2,cavo2a,cavo3,cavo4,cavo5,cavo6}.

\begin{figure}
\epsfxsize=10cm
\centerline{\epsffile{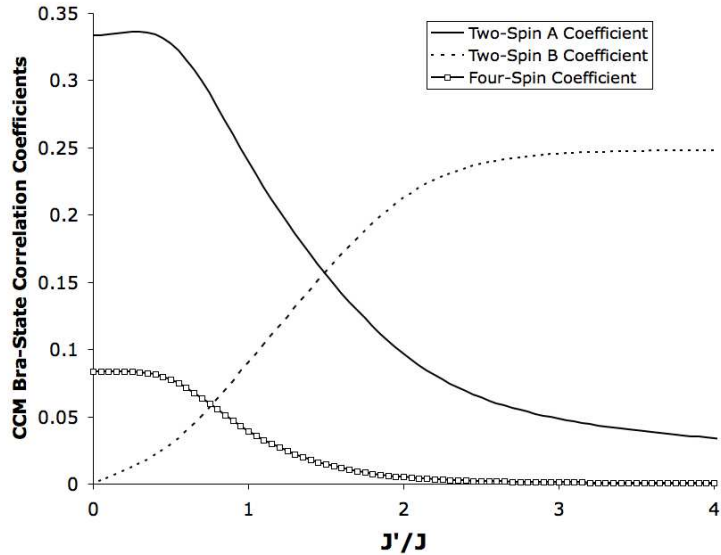}}
\caption{CCM results at the LSUB8 level of approximation for the ground-state nearest-neighbor
two-spin $A$ and $B$ bra-state coefficients ($\tilde {\cal S}_2^a$ and  $\tilde {\cal S}_2^b$)
and  the plaquette four-spin bra-state ($\tilde {\cal S}_4^p$)
bra-state correlation coefficients of the spin-half antiferromagnet on the
CAVO lattice.}
\label{cavo_bra_coeffs}
\end{figure}

The four-site plaquettes in the unit cell become decoupled only in the limit
$J'/J=0$. 
 The ground state is a product of such four-site plaquette singlets 
in this limit. 
To model such a state using the CCM we start again from the N\'eel model state; 
namely, a state in which the spins on nearest-neighbor sites are antiparallel. 
 
To create an exact plaquette-singlet product VBC ground state at $J'/J=0$ 
using the CCM we have to adjust
the 
nearest-neighbor correlation coefficients ${\cal S}_2^a$ and  ${\cal S}_2^b$
and a single four-body plaquette correlation  
coefficient ${\cal S}_4^p$ containing all four sites properly. 
(Note that  ${\cal S}_2^a$ represents those ket-state coefficients for 
the nearest-neighbor two-body cluster connecting sites on a plaquette 
indicated by the solid lines in Fig. \ref{cavo_unit_cell}, 
whereas ${\cal S}_2^b$ represents those ket-state coefficients 
for the nearest-neighbor two-body cluster connecting sites 
on a dimer indicated by the dotted lines in the same figure. 
The coefficient ${\cal S}_4^p$ represents those ket-state coefficients for 
the four-body cluster corresponding to a plaquette 
indicated by the solid lines in Fig. \ref{cavo_unit_cell}.)
Indeed, it is easy to show that setting the ket-state correlation 
coefficients ${\cal S}_2^a$ and ${\cal S}_4^p$ to a value of 0.5 and all other ket-state correlation
coefficients (including ${\cal S}_2^b$) to zero the  
plaquette-singlet
product
VBC state is obtained exactly, see
Fig.~\ref{cavo_ket_coeffs}.
 We are able to track  this plaquette solution 
as $J'/J$ is increased away from the 
 point $J'/J=0$ where it is exact. Furthermore, we are also able 
to reproduce exactly the  dimer-singlet product ground state 
in the limit 
$J'/J \rightarrow \infty$. In this limit, the nearest-neighbor 
ket-state correlation coefficient ${\cal S}_2^b$ on 
the dimer bonds (dotted lines in Fig.
\ref{cavo_unit_cell}) has a value of one 
and all other coefficients (e.g., ${\cal S}_2^a$ and
${\cal S}_4^p$) are zero, see
Fig.~\ref{cavo_ket_coeffs}. 
This solution is produced automatically when we track the 
solution (outlined above) from $J'/J=0$, and so our CCM Ansatz produces accurate results in all phases of this model. 
The corresponding bra-state correlation coefficients $ \tilde {\cal S}_2^a$, $\tilde {\cal S}_2^b$
and $\tilde {\cal S}_4^p$
behave smoothly in
the entire range  of $J'/J$, see Fig.~\ref{cavo_bra_coeffs}.

\begin{figure}
\epsfxsize=11cm
\centerline{\epsffile{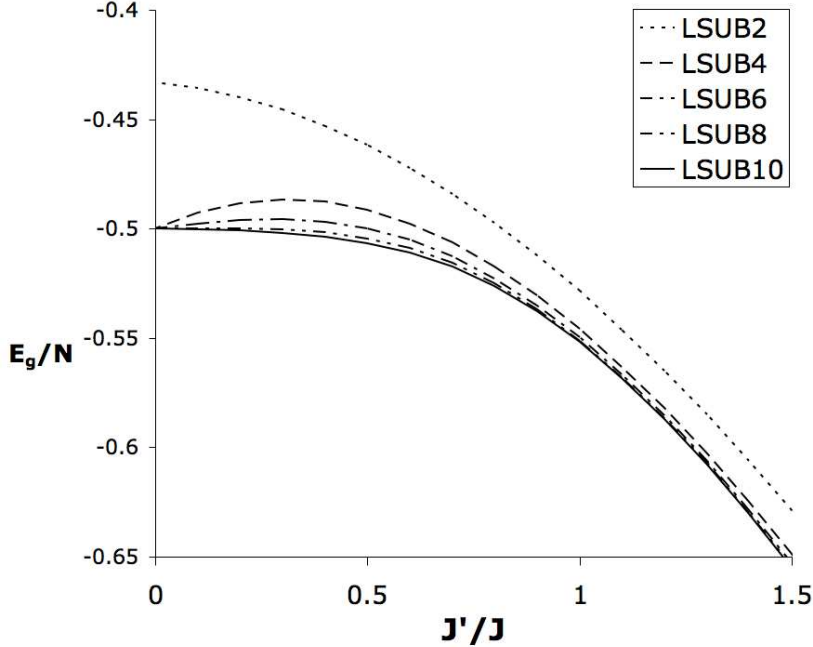}}
\caption{CCM results for the ground-state energy of the $J$--$J'$ 
Heisenberg antiferromagnet on the CAVO lattice (with $J=1$).}
\label{cavo_energies}
\end{figure}

Results for the ground-state energy are shown in Fig. \ref{cavo_energies}. We see that the exact ground-state energy at 
$J'/J=0$ is reproduced for LSUB$m$ levels of approximation with $m \ge 4$, as expected. At $J'/J=1$, 
we reproduce a previous result \cite{ccm26} using the N\'eel model state for this CAVO lattice. At all values of $J'/J$, 
the LSUB$m$ results are seen to converge rapidly with increasing levels of LSUB$m$ approximation. We note that the system should again decouple into dimers 
as $J'/J \rightarrow \infty$ and the correct ground-state energy ($-0.375J'$) is reproduced in this limit. 
The CCM provides excellent results for the ground-state energy for all values of
$J'/J$.
\begin{figure}
\epsfxsize=11cm
%\centerline{\epsffile{cavo_mag.ps}}
\centerline{\epsffile{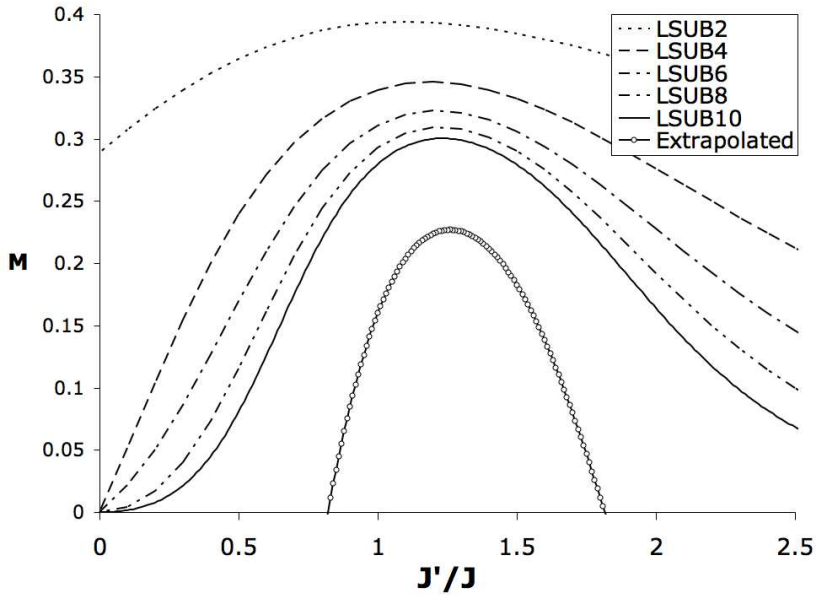}}
\caption{CCM results for the ground-state sublattice magnetization $M$
of the $J$--$J'$ 
Heisenberg antiferromagnet on the CAVO lattice.}
\label{cavo_mag}
\end{figure}

The results for the sublattice magnetization are shown in Fig. \ref{cavo_mag}. 
 We extrapolate the raw LSUB$m$ data to the limit $m\to\infty$ in order
to determine the quantum phase transition points where the magnetic N\'eel 
long-range order vanishes.   
An appropriate extrapolation rule for the magnetic order parameter 
for systems showing a ground-state order-disorder transition
is \cite{ccm30,ccm31,ccm32,ccm33,ccm34,ccm35}
$ M(n)=b_0+b_1(1/n)^{1/2}+b_2(1/n)^{3/2}$,
%and extrapolation 2 $ M(n)=c_0+c_1(1/n)^{\nu}$,
where we use LSUB$m$ results with $m=\{4,6,8,10\}$.
The results for these 
quantum critical points are shown in Table \ref{tab2}. 
Again, we note that these results indicate that the 
N\'eel order persists over a finite range around $J'/J=1$ for the 
of the $J$--$J'$ 
Heisenberg antiferromagnet on the CAVO lattice.
Finally, these results show that the CCM can handle  plaquette VBC 
ordering as easily as dimer VBC ordering  or the usual N\'eel ordering. 
Furthermore, we have demonstrated  that  quantum critical points can be 
determined by using the CCM to high orders of LSUB$m$ approximation. 

\begin{table}[t]
\caption{Results for the quantum critical points
 of the $J$--$J'$ Heisenberg antiferromagnet on the CAVO lattice.}
\begin{center}
{\footnotesize
\begin{tabular}{|l|c|c|}                                                                                      
\hline\noalign{\smallskip}
Method									&$J'_{c_1}/J$    &$J'_{c_2}/J$            	\\ 
\noalign{\smallskip}\hline\noalign{\smallskip}
CCM Extrapolation ($m=\{4,6,8,10\}$)		&0.82                    &1.82       	 \\
%CCM Extrapolation 2 ($m=\{4,6,8,10\}$)		&0.96                    &1.64       	 \\
QMC  \cite{cavo2,cavo2a}                                              &0.939	            &1.68$\pm$0.14 	\\
Cumulant Series Expansions  \cite{cavo4}		&0.9		          	&1.7		\\ 
Non-Linear Spin-Wave Theory  \cite{cavo6}		&0.90			&1.6		\\
Schwinger-Boson Mean-Field Theory  \cite{cavo5}	&0.6				&2.4		\\ 
\noalign{\smallskip}\hline
\end{tabular}
}
\end{center}
\label{tab2}
\end{table}

\section{Conclusions}
\label{sec:4}
  
We have shown in this article that we can easily form dimer and 
plaquette VBC ground states using the CCM with  
 independent-spin product model states. 
We have investigated a number of cases in which the ground state 
was a product of 
localized dimer or plaquette singlets. 
Firstly, we considered the spin-half $J_1$--$J_2$ model for the linear chain. 
We showed that we are able to reproduce exactly the 
dimerized ground state at $J_2/J_1=0.5$. 
 Interestingly, 
a symmetry-breaking dimerized CCM solution
is observed for $J_2/J_1<0.5$, which only becomes equal to the 
usual (`N\'eel-type') solution that conserves 
the lattice symmetry
at a 
CCM critical point $J_2/J_1|_{c_1}$. 
Results 
for the bra state correlation coefficients diverged at this point also. 
We took this to indicate the onset of the dimerized 
ground-state phase that breaks the translational lattice symmetry. 
We found that the dimerized phase extends over a finite range 
of values of $J_2/J_1$ both above and below 0.5. 
Results for the ground-state energy for the 
dimerized CCM solution were found to agree extremely well with the results of exact diagonalizations for $N=28$ and $N=32$ 
chains in the dimerized regime. The change from the 
 usual (`N\'eel-type') solution 
to the dimerized solution was also observed in the 
behavior of the sublattice magnetization.

We then considered the Shastry-Sutherland model and we demonstrated that the CCM can span the correct ground states in both the 
N\'eel and the dimerized phases. We found a CCM critical point for the 
dimerized solution that is ``driven'' by the CCM bra state alone. 
Furthermore, the position of this critical point is in good agreement with the known value for the phase transition point of this model. Results for the sublattice magnetization (that ought to be zero in the dimerized phase) were found to be generally good in the dimerized regime. For example, results for the sublattice magnetization at the LSUB6 level of approximation were found to be $|M| < 10^{-2}$ for $J_2/J_1>2$.  However, the sublattice magnetization was also found 
to be negative and to diverge at the critical point of the bra-state 
equations.

Finally, we considered a  spin-half 
Heisenberg antiferromagnet with nearest-neighbor bonds with respect 
to an underlying lattice that corresponds to that of the magnetic material CaV$_4$O$_9$. 
The four nearest-neighbor bonds that connect sites on 
a four-site plaquette have a bond strength given by $J$ 
and the nearest-neighbor bonds that connect sites belonging to 
different plaquettes (dimer bonds)
have a bond strength given 
by $J'$. The exact plaquette-singlet ground state 
at $J'/J=0$ and the 
exact dimer-singlet ground state  as $J'/J \rightarrow \infty$ were both 
reproduced exactly using the CCM with the same choice of N\'eel model state. 
We found that the CCM can 
provide precise results for the ground-state energy over all intervening 
values of the parameter $J'/J$.  
 Results for the sublattice magnetization were presented, 
and these results indicated that  
the N\'eel-ordered regime persists over a finite range of values of $J'/J$ 
around the point $J'/J=1$.
For large (and small) values of $J'/J$, the N\'eel long-range order is 
destroyed by local singlet formation on dimers (and on plaquettes). 
Extrapolations of LSUB$m$ data suggest that the 
N\'eel-order regime extends over the range $0.82 < J'/J < 1.82$. 
%using one extrapolation procedure and $ 0.96 < J'/J < 1.64$ using
%another. 
These results were found to be in fairly reasonable agreement 
with quantum Monte Carlo results for this model \cite{cavo2,cavo2a}.
However, a discussion of the accuracy of phase transition points estimated using
the CCM is beyond the scope of this article and so will form the contents of 
another article, although we note here that we believe also that higher 
orders of LSUB$m$ approximation would  provide closer agreement.

As noted above, the CCM is one of the most powerful and most widely 
applied techniques of quantum many-body theory. One of the reasons
for this success is based on the fact that the CCM allows the 
inclusion of multi-particle correlations into the ground- and excited-state 
wave functions in a controlled and systematic manner. The range of 
applicability of the CCM  to lattice quantum spin systems has been greatly 
extended previously by the creation of efficient and powerful high-order 
computer codes for  independent-spin product (e.g. N\'eel) model states. 
These codes are simple to use and they are generally accurate in 
practical applications. Furthermore, they are extremely flexible in terms of 
defining and solving new spin problems.  

Previously however, non-classical orderings (such as local singlet formation) have also been considered using the CCM 
by employing non-N\'eel model states. This typically involved the direct use of products of, 
e.g.,  local dimer singlets,  in the model state. 
However, this approach required a whole new matrix-operator formalism to be created for each 
new problem  \cite{ccm6}. 
This is usually tedious and time-consuming, although it is normally straightforward 
mathematically. More importantly, however, the existing high-order CCM codes would then 
need to be amended extensively 
also in order to implement the new matrix algebra for each new problem. Here we have presented a much simpler and more universal approach that 
combines exact solutions  for dimer or plaquette VBC product ground states 
with the computational implementation described in Refs.
 \cite{ccm8,ccm12,ccm15,ccm20,ccm23,ccm} based on 
 independent-spin product model states. 

One seeming shortcoming of this new approach was found to be that the ket state can be an exact representation of the true ground state, whereas the bra state might 
not be at the same level of LSUB$m$ approximation.  This is due to the simple fact that the NCCM parametrizations of the ket and bra wave functions 
 are not manifestly Hermitian conjugates of each other.  This meant, for example, that the exact ground-state energy of the Shastry-Sutherland model in the dimerized phase was reproduced, since it 
 is calculated from the ket-state correlation coefficients alone, 
 whereas the exact sublattice magnetization (known to be zero in this regime) was not,
 since its calculation requires the use of the bra-state correlation coefficients as well as the
 ket-state coefficients. 
 We speculate that this problem might be overcome by employing the extended coupled cluster method (ECCM) that contains an 
exponentiated form of the bra-state correlation operator in the (ground) bra state. We believe that it is possible to construct a similar exact bra-state solution for the ECCM  as we 
have constructed here for the ket state using the NCCM based on  
 independent-spin product model states.

As noted above, an alternative approach is to utilize model states that are 
formed directly from products of local dimer or 
plaquette states (even for the NCCM). In this case, the ket and bra state 
can become Hermitian conjugates (trivially) when the model state is the exact 
ground state. Indeed, in this case, all of the CCM ket- 
and bra-state correlation coefficients become zero. However, even here, we 
note that the lack of manifest Hermiticity is a general feature of the CCM, 
i.e., one that can exist even for results generated by calculations based on 
 such valence-bond model states. Indeed, in those cases for which the model state is not (trivially) the ground state, the bra and ket states are again not constrained manifestly to be Hermitian conjugates at any given level of LSUB$m$ approximation.

In conclusion, this new approach for dimer- and plaquette-ordered ground 
states is flexible, simple to implement, and very powerful. 
Lattices of arbitrary complexity can also be treated using this new method. 
Furthermore, this approach is simple because we are using  
generally  independent-spin product  model states 
derived from classical ground states. 
Indeed, this is far simpler than the alternative of creating a whole 
new matrix formalism for each new model state formed from products of 
localized states. Finally, this approach is powerful because the 
high-order codes based on  independent-spin product model states,  
which have been employed previously with great success, may be used directly in order to 
simulate the properties of these non-N\'eel states. The useful LSUB$m$ and 
SUB$n$ approximation schemes devised for the N\'eel model states may be used 
directly also. The results of the LSUB$m$ scheme may be extrapolated 
easily to the limit $m \rightarrow \infty$ using 
existing `heuristic' extrapolation schemes.  The results presented here 
offer a great enhancement to the range of applicability of the CCM for 
lattice quantum spin 
systems that demonstrate `novel states' of quantum order. 

\begin{acknowledgements}
One of us (DJJF) gratefully acknowledges support for the research presented 
here from the European Science Foundation 
(Research Network Programme: Highly Frustrated Magnetism).
 Two of us (RZ and JR) thank the DFG for support(project RI615/16-1).

\end{acknowledgements}



\end{document}